\documentclass[conference]{IEEEtran}
\IEEEoverridecommandlockouts
\usepackage{color}
\usepackage{tabularray}
\usepackage[table,xcdraw]{xcolor}
\usepackage{dblfloatfix}
\usepackage{cite}
\usepackage{multicol}
\usepackage{diagbox}
\usepackage{fancyhdr}
\fancypagestyle{firstpage}{%
  \lhead{This work has been accepted for publication in the IEEE ICMLCN 2024 Conference.}
}

\usepackage{adjustbox}
\usepackage{framed}
\usepackage{tabularray}
\usepackage{amssymb}
\usepackage{amsfonts}
\usepackage{amsmath}
\usepackage[T1]{fontenc}
\usepackage[utf8]{inputenc}
\usepackage[english]{babel}
\usepackage{pifont}
\usepackage{booktabs}
\usepackage{amsmath}
\usepackage{mathtools}
\usepackage{ellipsis}
\usepackage{textcomp}
\usepackage{cite}
\usepackage{subfigure}
\usepackage{epsfig}
\usepackage{epstopdf}
\definecolor{Mycolor1}{HTML}{ff0000}
\definecolor{Mycolor2}{HTML}{13beb8}
\definecolor{Mycolor3}{HTML}{ffea00}
\definecolor{Mycolor4}{HTML}{4dd0e1}
\definecolor{Mycolor5}{HTML}{ab47bc}
\definecolor{Mycolor6}{HTML}{ffa726}
\definecolor{Mycolor7}{HTML}{bcaaa4}
\definecolor{Mycolor8}{HTML}{43a047}
\definecolor{Mycolor9}{HTML}{81d4fa}
\definecolor{Mycolor10}{HTML}{999900}
\definecolor{Mycolor11}{HTML}{FFEBCC}
\definecolor{Mycolor12}{HTML}{D1E9E2}
\usepackage{algorithm}%\usepackage{algpseudocode}
\usepackage{algpseudocode}
\usepackage{epstopdf}
\usepackage{psfrag}
\usepackage{grffile}
\usepackage{amsthm}
\usepackage{caption}

\theoremstyle{definition}

\theoremstyle{remark}

\theoremstyle{plain}

\newcommand{\RNum}[1]{\uppercase\expandafter{\romannumeral #1\relax}}
\usepackage{setspace}

\def\BibTeX{{\rm B\kern-.05em{\sc i\kern-.025em b}\kern-.08em
    T\kern-.1667em\lower.7ex\hbox{E}\kern-.125emX}}
\begin{document}

\title{ML Framework for Wireless MAC Protocol Design \\
 \vspace{-0.15cm}
}
\author{\IEEEauthorblockN{ Navid Keshtiarast and Marina Petrova}
\IEEEauthorblockA{Mobile Communications and Computing, RWTH Aachen University, Aachen, Germany \\
Email: \{navid.keshtiarast, petrova\}@mcc.rwth-aachen.de}
\vspace{-0.7cm}
}
\maketitle
\thispagestyle{firstpage}
\begin{abstract}
   Adaptivity, reconfigurability and intelligence are key features of the next-generation wireless networks to meet the increasingly diverse quality of service (QoS) requirements of the future applications. Conventional protocol designs, however, struggle to provide flexibility and agility to changing radio environments, traffic types and different user service requirements.  In this paper, we explore the potential of deep reinforcement learning (DRL), in particular Proximal Policy Optimization (PPO), to design and configure intelligent and application-specific medium access control (MAC) protocols. We propose a framework that enables the addition, removal, or modification of protocol features to meet individual application needs. The DRL channel access policy design empowers the protocol to adapt and optimize in accordance with the network and radio
 environment. Through extensive simulations, we demonstrate the superior performance of the learned protocols over legacy \mbox{IEEE 802.11ac}  in terms of throughput and latency.
   %Our work presents a promising path towards satisfying the QoS requirements of future wireless applications by incorporating intelligence and adaptability into the protocol design.
\end{abstract}
%\vspace{-0.15cm}
\section{Introduction}
\label{sec:Introduction}
%\vspace{-0.05cm}
%Define the problem:
% The next-generation mobile technology 
% To satisfy the QoS for all future applications
% %%% ABI research
% since with the increasing diversity of applications and the widespread use of communication devices everywhere, human-designed protocols do not satisfy the flexibility and adaptability that different environments and diversified services and applications require.
% %plasticity
% Need a framework to create efficient protocols on demand for different applications. For example, on-demand converts the MAC protocol of the device with just communication functionality to a device with both sensing and communication functionality.
% Also, Future medium access protocols should be designed to allow learning in the environment to make informed design choices.[5G Americas]. 
% where we want to go:
% The future ai-based protocol should provide optimised network performance ranging from a small residential scenario with a simplistic environment to a big factory with complex environments. (Navid)
% Using deep neural networks in the system will help the system to adapt even to unknown environments or new traffic patterns.
% 1)make more efficient use of the spectrum
% traditional process of algorithm design 
%and channel access policies that fluently transition from contention- to schedule-based depending on the use case and environment 
%Flow: Why centralized, Why WLAN, Why RL
\subsection{Motivation}
One of the major challenges of next-generation wireless systems is to meet the diverse and demanding quality of service (QoS) requirements of future immersive applications. With the rise of applications beyond the classical voice and data (XR, digital twinning, autonomous driving to name a few), and the widespread use of communication devices everywhere, the current
%human-designed
existing protocols will be challenged to fulfil the flexibility and adaptability that different environments and diversified services and applications require. %To address this challenge,
At the center of these challenges lies the WLAN medium access control (MAC), with its key role in ensuring efficient use of the shared communication medium. However, this critical role comes with complexities, as the extensive number of parameters and their intricate inter-dependencies in the MAC layer pose significant challenges when employing conventional methods for optimizing and fine-tuning protocols, particularly within dense deployment scenarios\cite{Falko}. This complexity underscores the necessity for a paradigm shift in tuning parameters to meet the evolving demands of future wireless networks. To tackle these evolving demands, the adoption of artificial intelligence (AI) and machine learning (ML) techniques has gained a lot of interest among researchers in industry and academia. Also, the IEEE 802.11 AI/ML Topic Interest Group (TIG) has been founded to advance WI-Fi 8 and beyond by making AI/ML-driven protocols feasible\cite{giordano2023wifi}\cite{Carlos}.

In this paper, we aim to answer the following question: 
\textit{How can wireless MAC protocols be redesigned to provide the necessary flexibility and adaptability with the help of ML methods, in particular reinforcement learning for different deployment environments, diverse services and applications?}
%\vspace{-0.12cm}
\subsection{Related work}
%\vspace{-0.05cm}
In recent years, artificial intelligence (AI) and machine learning (ML) techniques have gained significant attention as promising approaches to tackle the tuning challenges, such as rate adaptation\cite{rate_Infocom} or optimizing MAC parameter in wireless communication systems\cite{Szymon_WCNC}. These efforts optimize mostly one or two parameters within the communication protocols. A comprehensive survey of studies that employ ML methods to enhance Wi-Fi technologies across different protocol layers can be found in \cite{Falko}. In contrast to using AI/ ML for only tuning one or two parameters, AI-based methods such as deep reinforcement learning (DRL) show promise to not only optimize the parameters of conventional algorithms but also generate new algorithms that can outperform conventional robust methods \cite{Fawzi2022}. In this regard, researchers recently started using the DRL to generate MAC design and configuration policies for cellular and wireless networks. In \cite{Jakob1} and \cite{Jean-Marie}, the authors use multi-agent reinforcement learning (MARL) to generate a new MAC protocol for uplink communication of UE in cellular networks, where each node learns signalling policy through the exchange of control messages.

In our previous work\cite{Peng}, we decomposed several MAC protocols into their atomic functional blocks and designed policies for wiring the needed blocks into a protocol based on the network environment using simple logic. 
In this paper, we make a major step forward and adopt an DRL on-policy approach for rewiring the MAC functional blocks into a protocol, which brings intelligence to the networks by allowing the protocol to learn the policy directly from its interactions with the network environment and adjust its behaviour accordingly. 

Prior research is primarily focused on tuning specific MAC parameters with respect to one performance metric and within certain scenarios, often overlooking the intricate interactions between these parameters. Moreover, many studies lack comprehensive validation within realistic network simulations \cite{Pasandi}. To address these gaps, we designed a framework using the OMNeT++ simulation platform that enables comprehensive validation and verification of our machine learning-based MAC protocol. Unlike previous studies, our proposed ML-based framework has the capacity not only to optimize multiple MAC parameters concurrently but also to iteratively redesign a novel MAC protocol tailored to the environment. This is facilitated through the agent's access to an array of MAC functions and their parameters. 

To this end, the main contributions of this work are summarized as follows:
\subsection{Contribution}
\begin{itemize}
    \item Protocol Design Framework: This paper proposes a framework that serves to explore the possibility of using \mbox{IEEE 802.11ac}s MAC protocol components to create new protocols using the DRL method. We design the framework in OMNET++ to decompose \mbox{IEEE 802.11ac} into a set of blocks and add, remove or modify these blocks to create new, application-specific MAC protocols.
    \item We show how DRL could be used to design better performing, i.e., higher throughput and lower latency MAC protocols for the next generation of wireless networks. We demonstrate the possibility of generating new application-specific MAC protocols by leveraging the legacy IEEE 802 11 ac protocol in the reward calculation function. % for our learning framework. 
    \item Superior Performance: We show through extensive system-level simulations that due to its learning-based nature and the ability to interact with the environment and use different rewards, such as throughput in the learning procedure, the DLR-based synthesis protocol outperforms the legacy protocol across diverse scenarios. This indicates that our framework has the potential to enhance the QoS requirements for future applications.
    \item Public access to the framework's code: We provide public access to our framework code at \cite{Rec_mac}, which promotes a deeper understanding of our technical approach and encourages the reproduction and validation and extension of our work.
\end{itemize}
The remainder of the paper is organized as follows. In Section II, we discuss the related works. Section III provides the system model and the simulation setup. In Section IV, we model our AI-based MAC protocol by using the DRL method. Section V shows and discusses the performance evaluation of the designed protocol. In Section VI, we conclude the paper.
\section{System Setup}
\begin{figure}[h!]
\vspace{-0.25cm}
	\centering
	\includegraphics[width=0.50\textwidth,trim = 73mm 55.5mm 52mm 33mm,clip]{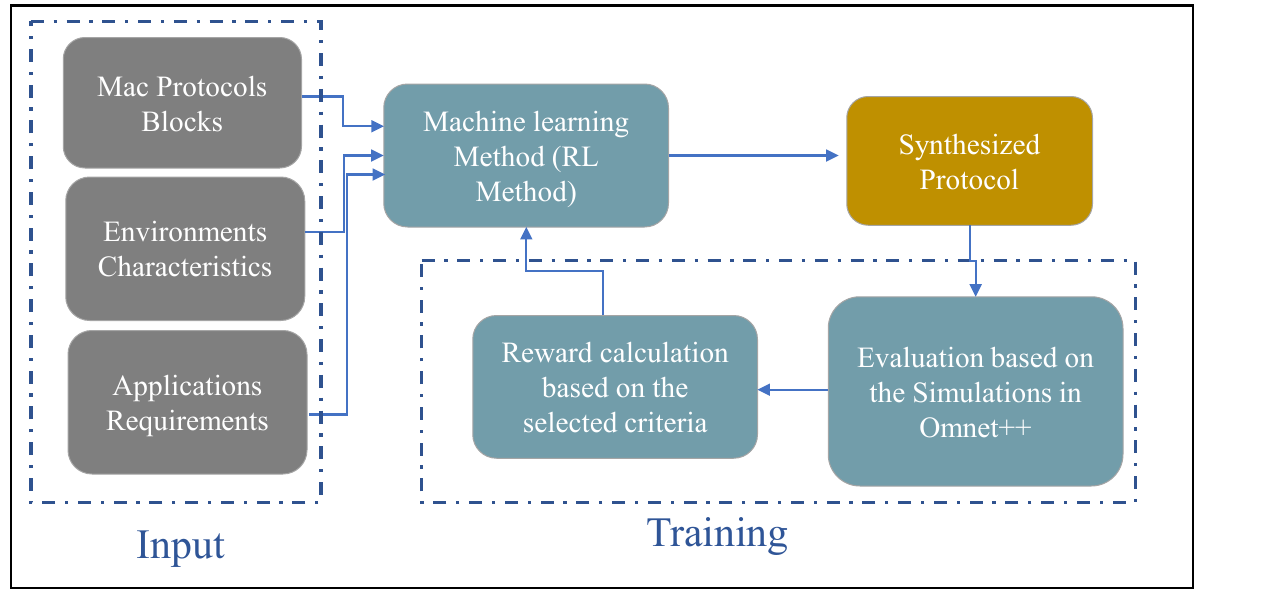}
	\caption{System architecture}
	\label{fig:System_model1}
 \vspace{-0.25cm}
\end{figure}

Figure \ref{fig:System_model1} illustrates the system architecture for composing intelligent, learning-based MAC protocol. 
We decompose a number of protocols that share components into a set of atomic blocks, which serve as the fundamental building blocks of our intelligent MAC protocol. We use ML-driven policies to connect or wire these components to create the best possible communication protocol for the given environment.
Within the machine learning block, a reward-based mechanism guides the learning process, continually updating the behaviour of the MAC protocol based on real-time network conditions. This adaptive behaviour adjustment leads to enhanced network performance and increased reliability.
As shown in Figure \ref{fig:System_model1}, we give the various MAC block functions and environment and application requirements as input to the ML framework. The ML block syntheses a new protocol, which is evaluated in our link-level simulator by a defined reward (which is selected based on the application requirement). When the reward is calculated, it sends feedback to our ML framework. This procedure is iterated until a protocol is built with the best set of building blocks for the current network/environment configuration.
% show the system architecture for composing intelligence MAC protocol. 
%As an example, we decouple the 802.11 ac DCF into a set of individual functions such as carrier sensing and backoff function etc.
To demonstrate the flexibility of our approach, we take the \mbox{IEEE 802.11ac} distributed coordination function (DCF) and decouple it into a set of individual functions such as carrier sensing and backoff, etc. We chose to focus on DCF since it forms the foundation of the \mbox{IEEE 802.11ac} MAC standards that are widely used, and it already contains the necessary building blocks for creating a new MAC protocol. It includes essential functions like sensing the channel and managing the wait time before transmitting. This allows the MAC functions to be used for the creation of new protocols as well as conventional standards like ALOHA and CSMA.

\begin{figure}[h!]
 \vspace{-0.25cm}
	\centering
	\includegraphics[width=0.50\textwidth,trim = 30mm 46mm 28mm 38mm,clip]{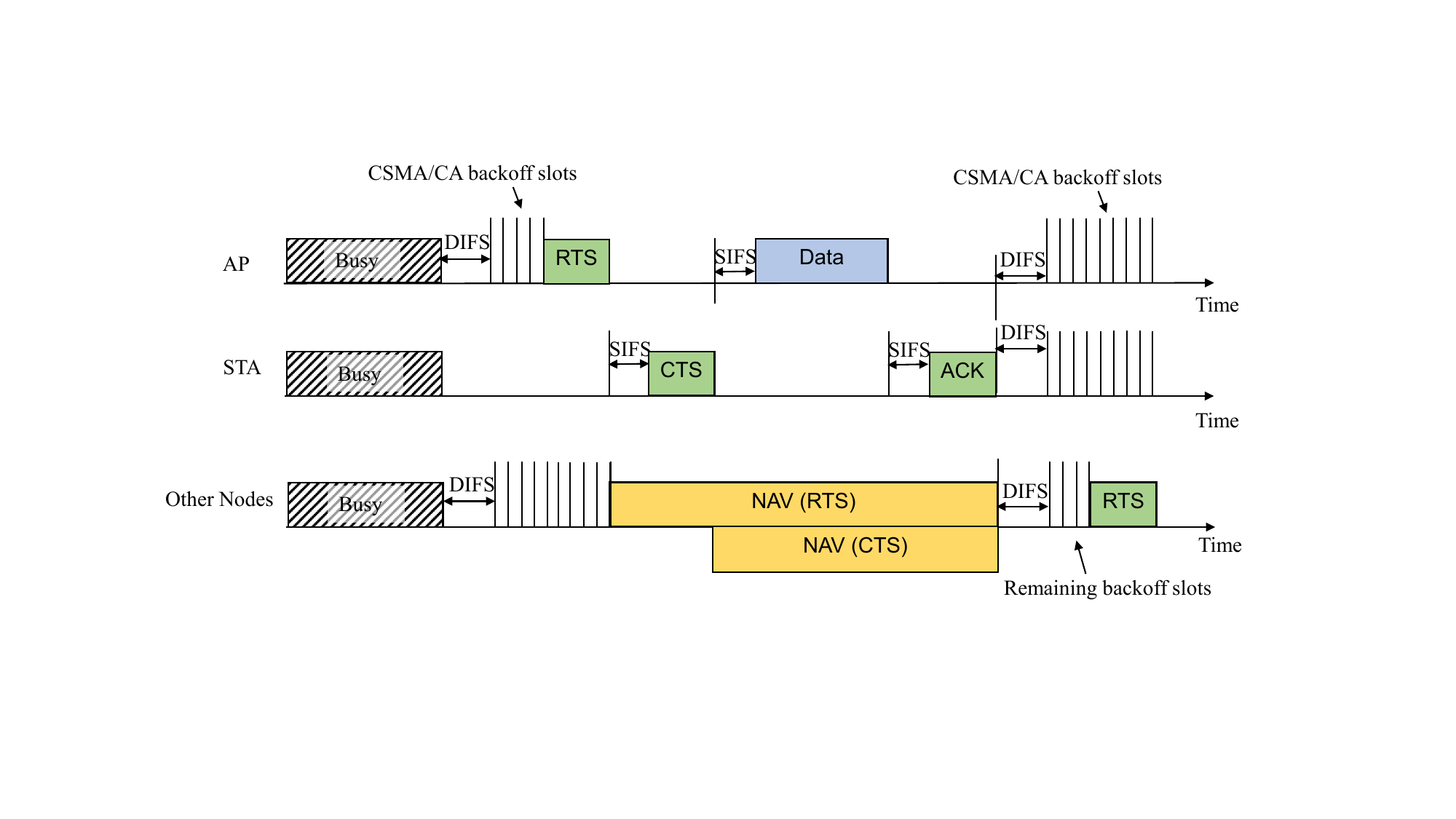}
	\caption{IEEE 802.11 DCF protocol with all MAC functions}
	\label{fig:DCF_}
 \vspace{-0.25cm}
\end{figure}

In Figure \ref{fig:DCF_}, we show the DCF, which is contention-based MAC in \mbox{IEEE 802.11ac} networks. DCF contains several functions, including carrier sensing (CS) and backoff, network allocation vector (NAV), and clear to send (CTS)/ request to send (RTS). The carrier sensing function is responsible for sensing the channel, and backoff defines a waiting time before transmission. The NAV is responsible for keeping the duration for which the channel is reserved by a frame transmission, and it prevents other nodes from transmitting during that time. RTS and CTS are the control frames transmitted by the sender and receiver, respectively, to trigger the NAV in surrounding nodes\cite{802.11ac}.

\begin{figure}[h!]
 \vspace{-0.15cm}
	\centering
	\includegraphics[width=0.5\textwidth,trim = 43mm 30mm 70mm 27mm,clip]{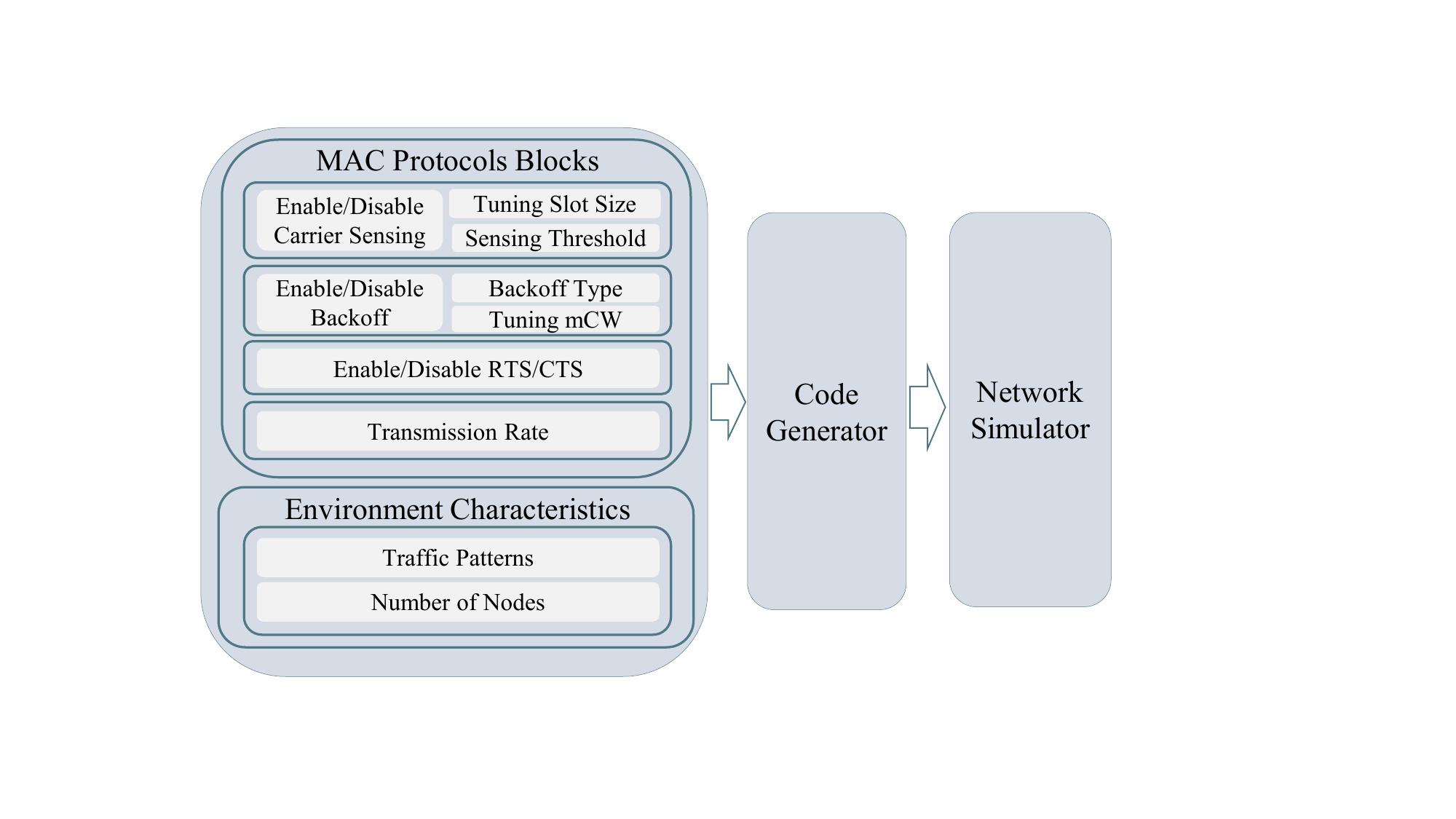}
	\caption{Simulation and Evaluation Setup}
	\label{fig:System_model2}
  \vspace{-0.25cm}
\end{figure}

Figure \ref{fig:System_model2} shows the simulation and evaluation setup. In the first block, the common MAC block functions of IEEE 802.11 are decomposed, enabling the activation or deactivation of individual blocks alongside parameter adjustments. For example, by disabling the carrier sensing block and backoff block, the MAC protocol converts to Aloha. In our framework, all decisions for excluding, including and tuning the parameter are made by the DRL agent.

%As environmental characteristics, we consider the traffic pattern and the number of surrounding parallel networks.
In addition to the MAC blocks, we also consider environmental characteristics such as traffic patterns and the presence of neighbouring parallel networks. These parameters are given in Table \ref{tab:Training Parameter}. Based on the output of the first block, the second block, i.e., code generator, creates the configuration file for the network simulator.
% \begin{figure}[h!]
% 	\centering
% 	\includegraphics[width=0.5\textwidth,trim = 55mm 55mm 70mm 27mm,clip]{pic/System2_modified_neu3.pdf}
% 	\caption{Simulation and Evaluation Setup}
% 	\label{fig:System_model2}
% \end{figure}

For conducting the simulations, we used the well-known \mbox{OMNeT++}\cite{omnet}, and INET \cite{INET} frameworks. We chose these frameworks due to their module-based architecture, which is compatible with our proposed concept. Within these frameworks, we made the MAC functions of \mbox{IEEE 802.11ac} modular, allowing the DRL agent to turn them on or off and tweak their settings as needed. Furthermore, in addition to the BEB, we also added two more backoff functions, i.e., exponential increase and exponential decrease (EIED) and constant backoff. 
\section{Deep reinforcement Learning}
\subsubsection{Objective: Maximize Mean throughput}
Our objective is to optimize the mean throughput across the entire network environment, considering all time steps and number of networks.
\begin{equation} \label{eqn0}
\max \frac{1}{T}\frac{1}{N} \sum_{t\in\{1,..,T\}, j\in\{1,...,N\}} U_j(t)
\end{equation}
Where $U$ is the aggregated downlink throughput per AP 
\subsubsection{Markov decision process (MDP)}
The agent selects an appropriate MAC policy for the current environment, which is applied to each node. The problem can be expressed as an MDP, which is defined by the tuple $(S, A, P, r,\gamma)$, where $S$ and $A$ represent the state and action spaces, respectively. $P$ is the distribution of transition probabilities between states, $r$ is the reward function, and $\gamma$ is the discount factor.

\textbf{Action $A$:} In each step, the agent will select the best possible MAC function and parameter for the current environment. We define action space as a vector with individual discrete space, i.e. multi discrete space. Each vector element represents one of the MAC functions or parameters. This action space allows each vector element to have a different number of choices. The corresponding action space $A$ can be represented as $A=(a_1,a_2,a_3,...a_7)$,  All possible vector elements for action space are summarised in Table \ref{tab:parameters}. For example, for the CTS/RTS, two actions are defined: CTS/RTS on or off. In the simulation, we use zero and one to show whether this function is off or on.
\begin{table}[h!]
\small
\centering
	\caption{The Action Space}
	\begin{tabular}{|p{0.3cm}|p{2.55cm}|p{3.45cm}|}
		\hline
		\textbf{} &  \textbf{Action parameter} & \textbf{Values} \\
		\hline
		$a_1$ &Carrier sensing & On or Off
		\\
		\hline
		$a_2$& Slottime&${5, 9, 20}$
		\\
		\hline
		$a_3$& Backoff type
		& {Off, EDID, BEB, Constant}   \\
		\hline
		$a_4$& Minimum CW
		&  ${15, 31, 63}$
        \\
		\hline
		$a_5$
		 & RTS/CTS& On or Off
		\\
		\hline
	    $a_7$& Datarate & $\{6.5, 13, 19.5, 26, 39,$\\&&$ 52, 58.5, 65, 78 \} Mbps$ 
		\\
		\hline
	\end{tabular}
	\label{tab:parameters}
  \vspace{-0.1cm}
\end{table}

\textbf{Observation space $o$}: In each time step, the observation space is given by $o=~ <<TR, NN, a>>$, where TR is the traffic characteristic, $N$ is the number of networks in the environment, and $a$ is the agents' action. Since the environment is entirely observable, state space and observation space are equivalent. The relevant parameters are presented in \mbox{Table \ref{tab:Training Parameter}}.

% \begin{table}[h!]
%  \vspace{-0.25cm}
% \small
% \centering
% 	\caption{Training Parameters}
% 	\begin{tabular}{|p{3.4cm}|p{3.4cm}|}
% 		\hline
% 		\textbf \textbf{Learning Rate} & 0.007\\
% 		\hline
% 		 Optimizer & Adam
% 		\\
% 		\hline
% 		Number of time slots, $T$ & 1000
% 		\\
% 		\hline
%         Number of training episode, $N_{eps}$ & 4000000
%         \\
%         \hline
% 		$\mbox{Clipping factor}$, $\epsilon$&0.2 \\
% 		\hline
%         $\mbox{Loss weighting} c1, c2 $,& 0.5, 0.02  \\
% 		\hline
%          $\mbox{Discount factor} \gamma$&0.99  \\
% 		\hline
% 		Policy & MLP (2 layers of 64) 
%         \\
% 		\hline
%         Mini batch size, $M$&64
%         \\
% 		\hline
% 	\end{tabular}
% 	\label{tab:Training Parameter}
%  \vspace{-0.25cm}
% \end{table}
% \begin{table}[h!]
%  \vspace{-0.15cm}
% \small
% \centering
% 	\caption{Environment Characteristic}
% 	\begin{tabular}{|p{4cm}|p{2.8cm}|}
% 		\hline
% 		 Number of networks (NN) & 1-30\\
% 		\hline
% 		 Frequency & 5 GHz
% 		\\
% 		\hline
%         Tx power & 23 dBm
%         \\
% 		\hline
% 		bandwidth & 20 MHz
% 		\\
% 		\hline
% 		Layout&$150\times150m^2$  \\
%         \hline
% 		Number of seeds for each action&$25$  \\
% 		\hline
% 		Traffic characteristic (TR): & $\lambda=20, 35, 50, 100,$\\Poisson Distribution with &${150, 200, 250, 300} $ \\ Arrival Rates $\lambda$ &\\
% 		\hline
% 	\end{tabular}
% 	\label{tab:Environment Parameter}
%  \vspace{-0.15cm}
% \end{table}
\begin{table}[h!]
 \vspace{-0.25cm}
\small
\centering
	\caption{Training Parameters}
	\begin{tabular}{|p{3.4cm}|p{3.4cm}|}
		\hline
		 Number of networks (NN) & 1-30\\
		\hline
		 Frequency & 5 GHz
		\\
		\hline
        Tx power & 23 dBm
        \\
		\hline
		bandwidth & 20 MHz
		\\
		\hline
		Layout&$150\times150m^2$  \\
        \hline
		Number of seeds for each action&$25$  \\
		\hline
		Traffic characteristic (TR): & $\lambda= 35, 50, 100, 200,$\\Poisson Distribution with &${300, \mbox{Saturated}} $ \\ Arrival Rates $\lambda$ &\\
		\hline
		\textbf \textbf{Learning Rate} & 0.007\\
		\hline
		 Optimizer & Adam
		\\
		\hline
		Number of time slots, $T$ & 1000
		\\
		\hline
        Number of training episode, $N_{eps}$ & 4000000
        \\
        \hline
		$\mbox{Clipping factor}$, $\epsilon$&0.2 \\
		\hline
        $\mbox{Loss weighting} c1, c2 $,& 0.5, 0.02  \\
		\hline
         $\mbox{Discount factor} \gamma$&0.99  \\
		\hline
		Policy & MLP (2 layers of 64) 
        \\
		\hline
        Mini batch size, $M$&64
        \\
		\hline
	\end{tabular}
	\label{tab:Training Parameter}
 \vspace{-0.15cm}
\end{table}

\textbf{Reward $R$:}
Depending on the requirements of the application, we are able to select as a reward either throughput or latency or any key performance indicators (KPIs) of the network. 
In this work, we use a throughput-based reward, which corresponds to our defined objective.
Since we want to design a protocol that surpasses the \mbox{IEEE 802.11ac},  we use the mean throughput of the entire network resulting from the simulation of the designed MAC protocol by the agent, and we subtract from it the mean throughput achieved from the same simulation scenario with the \mbox{IEEE 802.11ac} MAC protocol. We then define rewards as follows:
 \begin{equation} \label{eqn1}
 R =	\frac{\sum_{i=1}^N\big({x_i}-{y_i}\big)}{N}, 
 \end{equation}
where $x_i$ is the mean aggregated downlink throughput of $i_{th}$ network and $y_i$ is the mean aggregated downlink throughput of $i_{th}$ network when all networks use \mbox{IEEE 802.11ac} MAC protocol. 
In case the selected MAC blocks give output, we calculate the mean throughput of all networks.

\subsubsection{Proximal policy optimization (PPO) for designing MAC protocol}
Each episode is made up of $T$ consecutive time slots. The agent selects a set of MAC blocks for each time slot. We perform the simulations in our link-level simulator for the selected action in each time slot and collect the rewards. During training, we randomly sample different actions, i.e., different sets of MAC blocks, using the policy $\pi_\Theta(a|s)$. Here, $\pi_\Theta$ denotes the stochastic policy parameterized by $\Theta$, with the agent choosing the action $a$ in the state $s$. Our objective is to acquire an optimal policy $\pi_\Theta(a|s)$ that maximizes the cumulative reward in a defined scenario. 
To achieve this objective, we adopt the Proximal Policy Optimization (PPO) model, which follows the actor-critic approach. In this approach, we leverage two neural networks: the critic network represented by $V^{\pi}(s)$ responsible for estimating the value function, and the actor represented by $\mathcal{\pi}(a|s)$, responsible for estimating the policy. 

The PPO agent learns online, directly from the interaction with the environment. During these interactions, the agent gathers experience data with a size of $N$. Subsequently, this collected experience is utilized to update both the policy and value components. In the case of using the Policy gradient, the gradient estimator can be achieved by differentiating the objective function below.
\begin{align}
    &L^{\mathrm{PG}}(\Theta) = \mathbb{E}[\log \pi(a|s;\Theta) A))]. \label{eq:PPO1}
\end{align}
where $A = \mathcal{Q}^{\pi}(a|s) - V^{\pi}(s)$ is the estimated advantaged function. This advantage function is calculated by subtracting the value function ($V^{\pi}(s)$), which is the estimation of cumulative reward from the current state, from the discounted rewards ($\mathcal{Q}^{\pi}(a|s)$). The resulting advantage value reveals if our agent's action was better or worse than what was expected. 

Using Equation \ref{eq:PPO1} often leads to frequent policy updates, which have a negative effect on the training. To overcome this issue, Trust region policy optimization (TRPO)\cite{schulman2017trust} was introduced, which modified the objective function by incorporating the ratio of the probability of the current policy over the previous policy ($r(\Theta) = \pi_\Theta(a|s) /  \pi_{\Theta_{\mathrm{old}}}(a|s)$.) and solving a second-order optimization problem. Due to the complexity of this TRPO method, the PPO algorithm \cite{schulman2017proximal} was introduced with the following main objective function:
% \begin{equation}
%     \nabla_{\Theta} J(\Theta) = \nabla_{\Theta} \log \pi(a|s;\Theta) (\mathcal{Q}^{\pi}(a|s) - V^{\pi}(s)),
% \end{equation}
\begin{align}
    &L^{\mathrm{CLIP}}(\Theta) = \mathbb{E}[\min(r(\Theta) A, \mathrm{clip}(r, 1 - \epsilon, 1 + \epsilon) A)],
\end{align}
 where $\epsilon$ is used inside the clip function to limit the policy update, mitigating potential adverse effects on the agent's stability and learning process.

 The final objective function for agent training is obtained by adding two additional terms, namely, $L^{\mathrm{VF}}(\Theta)$ and entropy bonus term. $L^{\mathrm{VF}}(\Theta)$ is the baseline update from the critic network, which is equal to the squared difference between the estimated value $V_{\phi}(s)$ and the target value. The entropy bonus term makes it possible for the agent to have adequate exploration, and is defined as 
 %Thus, the objective function becomes as follows:
 \begin{equation} \label{eq:PPO33}
     S[\pi_\Theta(a_t|s_t)]=-\sum_a\pi_\Theta(a_t|s_t)\log \pi_\Theta(a_t|s_t),
     \vspace{-0.1cm}
\end{equation}
Thus, we can write the final loss function as follows:
 \begin{equation} \label{eq:PPO22}
     L^{}(\Theta)=L^{\mathrm{CLIP}}(\Theta) - c_1L^{\mathrm{VF}}(\Theta) + c_2S[\pi(.|s;\Theta)].
\end{equation}
where $c_1$ and $c_2$ are the hyper-parameters to give different weights to each term. Table \ref{tab:Training Parameter} shows the parameters used to train the agent. Algorithm \ref{Iterative_Algorithm} provides a summary of the training approach.
\begin{algorithm}[t]
\hspace*{\algorithmicindent} \textbf{Input:} 
 {All training parameters from Table \ref{tab:Training Parameter}\\}
 \hspace*{\algorithmicindent} \textbf{Output:}
 {$\pi_\Theta$, $V_\phi$ }
	\begin{algorithmic}[1]
	%\SetAlgoLined
    % \INPUTS \State All training parameters from Table \ref{tab:Training Parameter}
    \State Initialize the actor $\pi_\Theta(a|s)$ and critic $V_\phi$ with random parameters $\theta$ and $\phi$, respectively.
	\For{\mbox{iteration=1, 2,...$N_{eps}$}}
    %\For{{actor=1, 2,...N do}}
    \State Initialize the environment
    \State $j=0$
    \While{$ j< T$}
    \State {Generate an experience set of $N$ time steps by following MAC block policy $\pi_\Theta$ and collecting $(s_t,a_t,r_t)$}
    %In each time step, select an appropriate MAC block policy for the current environment by following $\pi_\Theta$ and  collecting $(s_t,a_t,r_t)$}
    %by following $\pi_\Theta$ and  collecting $(s_t,a_t,r_t)$
    \State Calculate the state values from the critic network 
    \State {Calculate the probability of action $\pi_\Theta(a_t|s_t)$ from actor network}
    \State {For each step calculate advantage function $A_t=\sum_{i=t}^L\gamma^{i-k}r_i-V_\phi(s_t)$ and return function $G_t=A_t+V_\phi(S_t)$}
    \State {Collect a subset of $M$ random sample (mini-batches) from the current set of experience}
    \State {Compute ratio $r_t(\theta)$and entropy loss $S[\pi(a_t|s_t)]$}
    \State {calculate value function loss  $L^{VF}=\frac{1}{2M}\sum_{i=1}^M(G_i-V_\phi(s_i))$ and surrogate objective $L^{\mathrm{CLIP}}=\frac{1}{M}\sum_{i=1}^M[\min(r(\Theta) A, \mathrm{clip}(r, 1 - \epsilon, 1 + \epsilon) A)]$}
    \State {Calculate the total loss $L(\Theta)=L^{\mathrm{CLIP}}(\Theta) - c_1L^{\mathrm{VF}}(\Theta) + c_2S[\pi_\Theta(a_t|s_t)]$ }
    \State {Update the actor and critic network parameters by minimizing the total loss}
    \State $j=j+N$
    \EndWhile
    %\State In each time step, select an appropriate MAC block policy for the current environment by following $\pi_\Theta$ and apply it on all nodes.
    %\State Compute the target value ($V^{target}$) and the advantage estimate (A) for each time step.
    %\State Perform gradient ascent to update the weights of the MAC selection policy.
     \EndFor   
		\caption{MAC Block Selection with Proximal Policy Optimization}
		\label{Iterative_Algorithm}
	\end{algorithmic}
\end{algorithm}

\begin{figure}[t]
	\centering
\includegraphics[width=0.4\textwidth,trim = 0mm 1mm 0mm 0mm,clip]{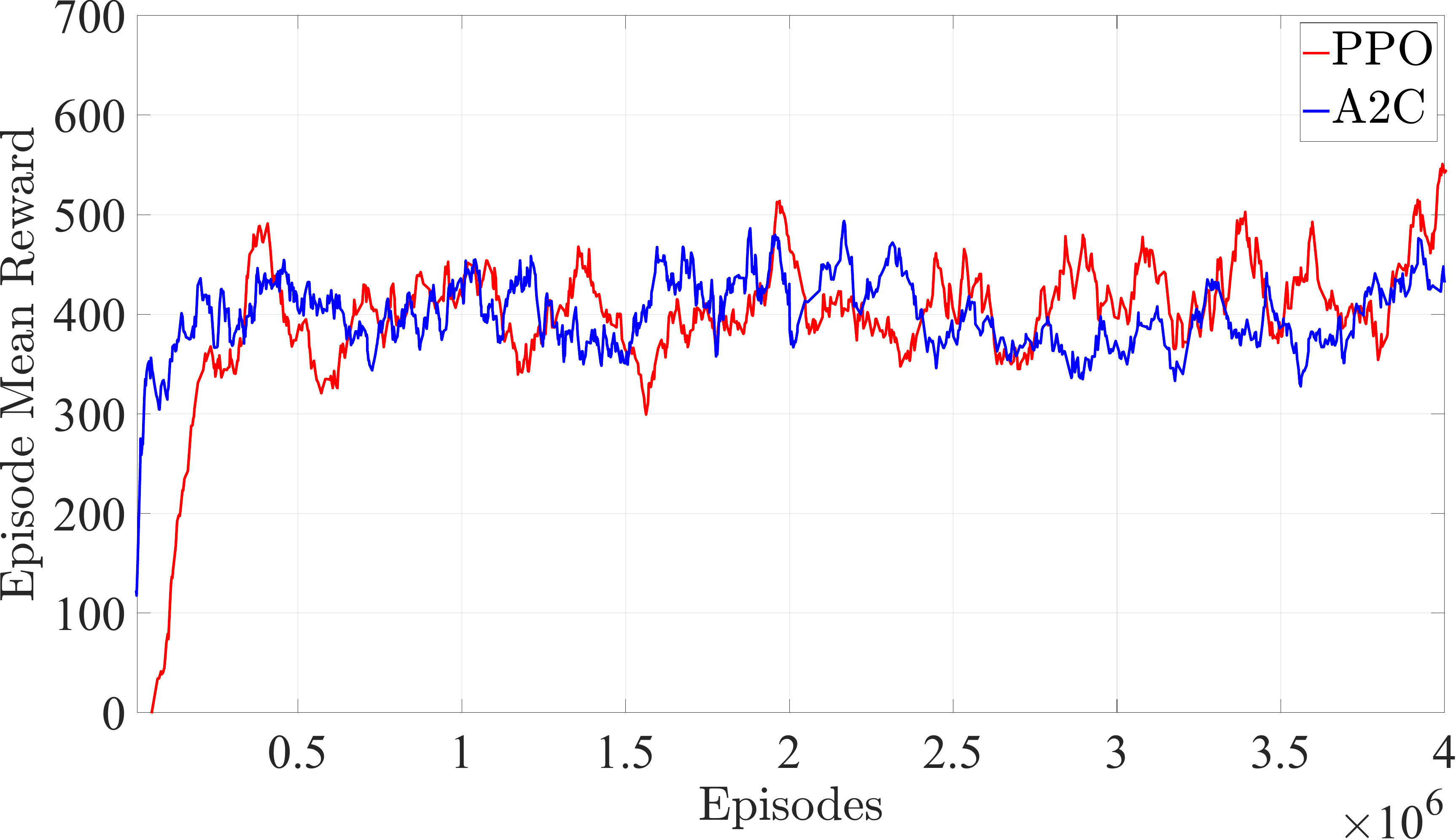}
	 \vspace{-0.2cm}
 \caption{Learning rate for A2C and PPO}
	\label{fig:Learning_rate}
 \vspace{-0.65cm}
\end{figure}

Figure \ref{fig:Learning_rate} illustrates the mean reward obtained by two DRL agents, i.e., PPO and A2C methods, during different episodes of training. The episode mean reward is relatively low and negative at the start of the training process, showing that the agent is performing poorly. As the training advances, the agent becomes more adept at navigating the environment, and the episode's reward begins to improve. The PPO learning curve reaches its highest value at the end of the training process, which shows the agent has learned is an effective policy and, in the final steps, consistently outperforms the A2C method.
\begin{figure*}[!ht]
%\hlin
% \resizebox{0.315\textwidth}{5cm}{%
%     \fbox{%
% \framebox[10cm][10cm]{}
\fbox{%
\parbox{0.313\textwidth}{%
	\centering
	\subfigure{
		\includegraphics[width=0.315\textwidth,trim = 0mm 0mm 0mm 0mm,clip]{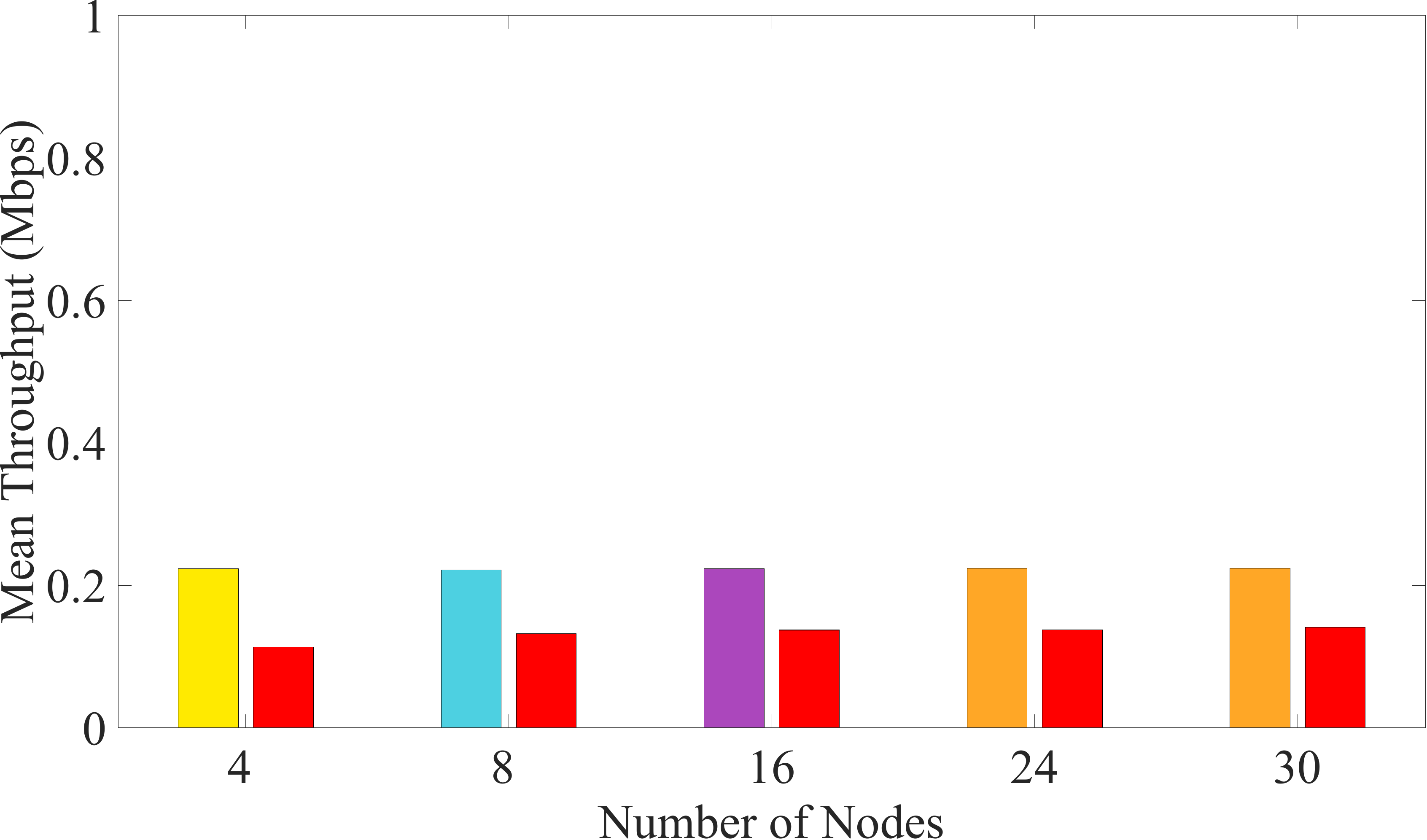}
		\label{TSUS_1}	
	}
 \addtocounter{subfigure}{-1}
 \subfigure[{20 packets/second average arrival rate}]{
		\includegraphics[width=0.315\textwidth,trim = 0mm 0mm 0mm 0mm,clip]{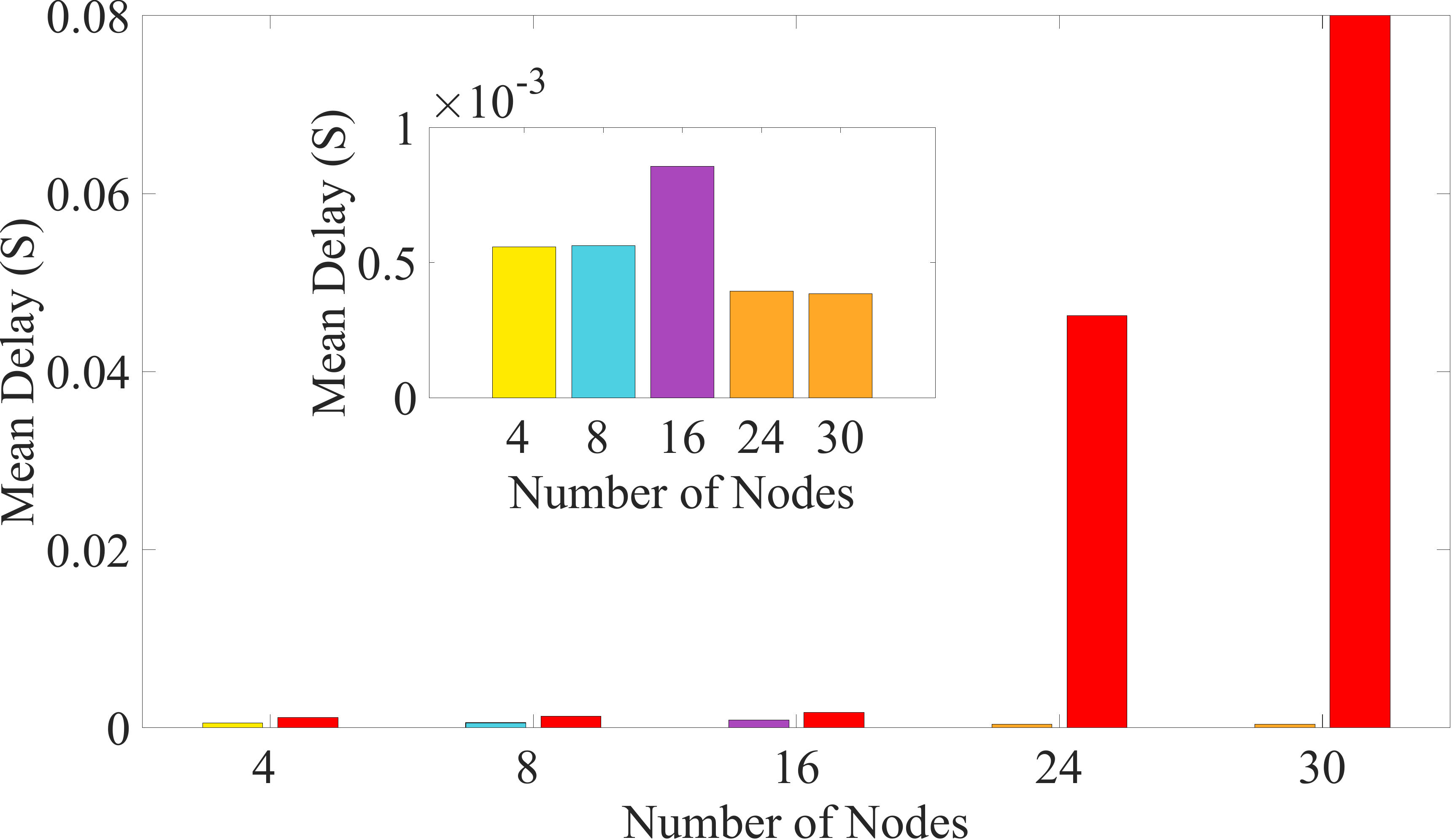}
		\label{TSUS_2_1}	
	}
 }
 }
 \fbox{%
\parbox{0.313\textwidth}{%
% }}
% \vrule
	\subfigure{
		\includegraphics[width=0.317\textwidth,trim = 0mm 0mm 0mm 0mm,clip]{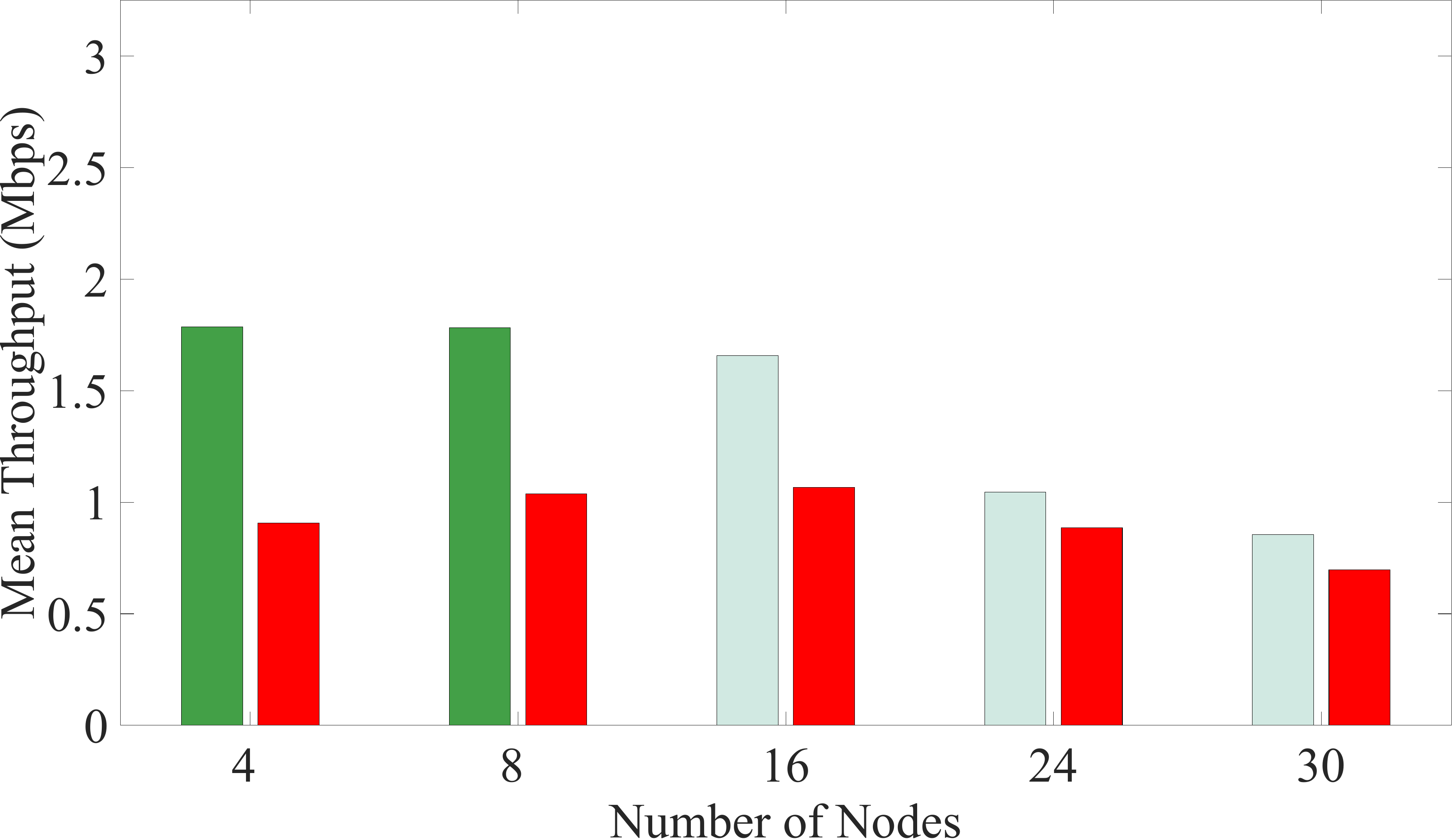}
		\label{TSUS_2}	
	}
  \addtocounter{subfigure}{-1}
 \subfigure[{150 packets/second average arrival rate.}]{
		\includegraphics[width=0.317\textwidth,trim = 0mm 0mm 0mm 0mm,clip]{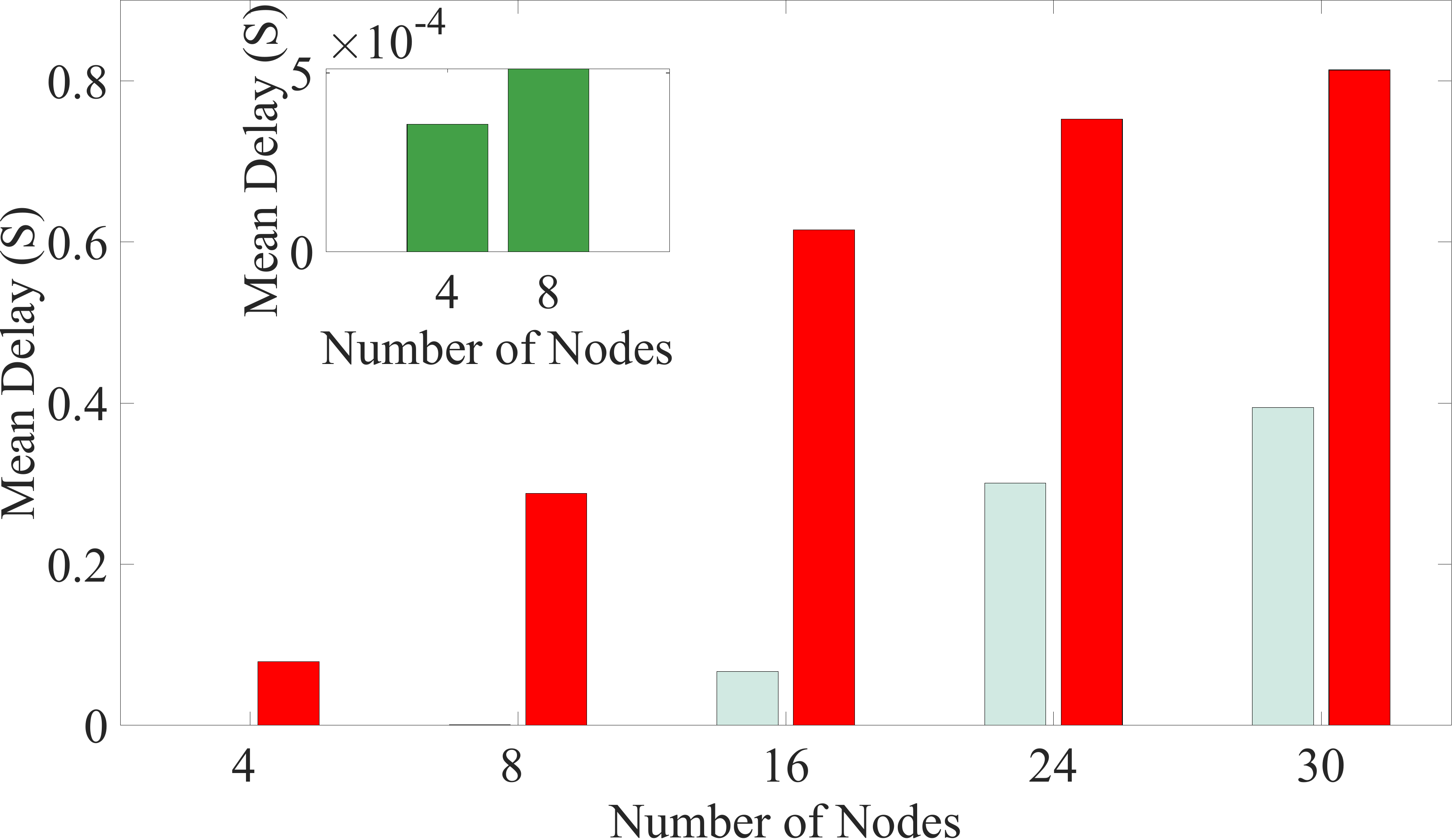}
		\label{TSUS_2_2}	
	}
 }
 }
% \vrule
 \fbox{%
\parbox{0.313\textwidth}{%
	\subfigure{
		\includegraphics[width=0.315\textwidth,trim = 0mm 0mm 0mm 0mm,clip]{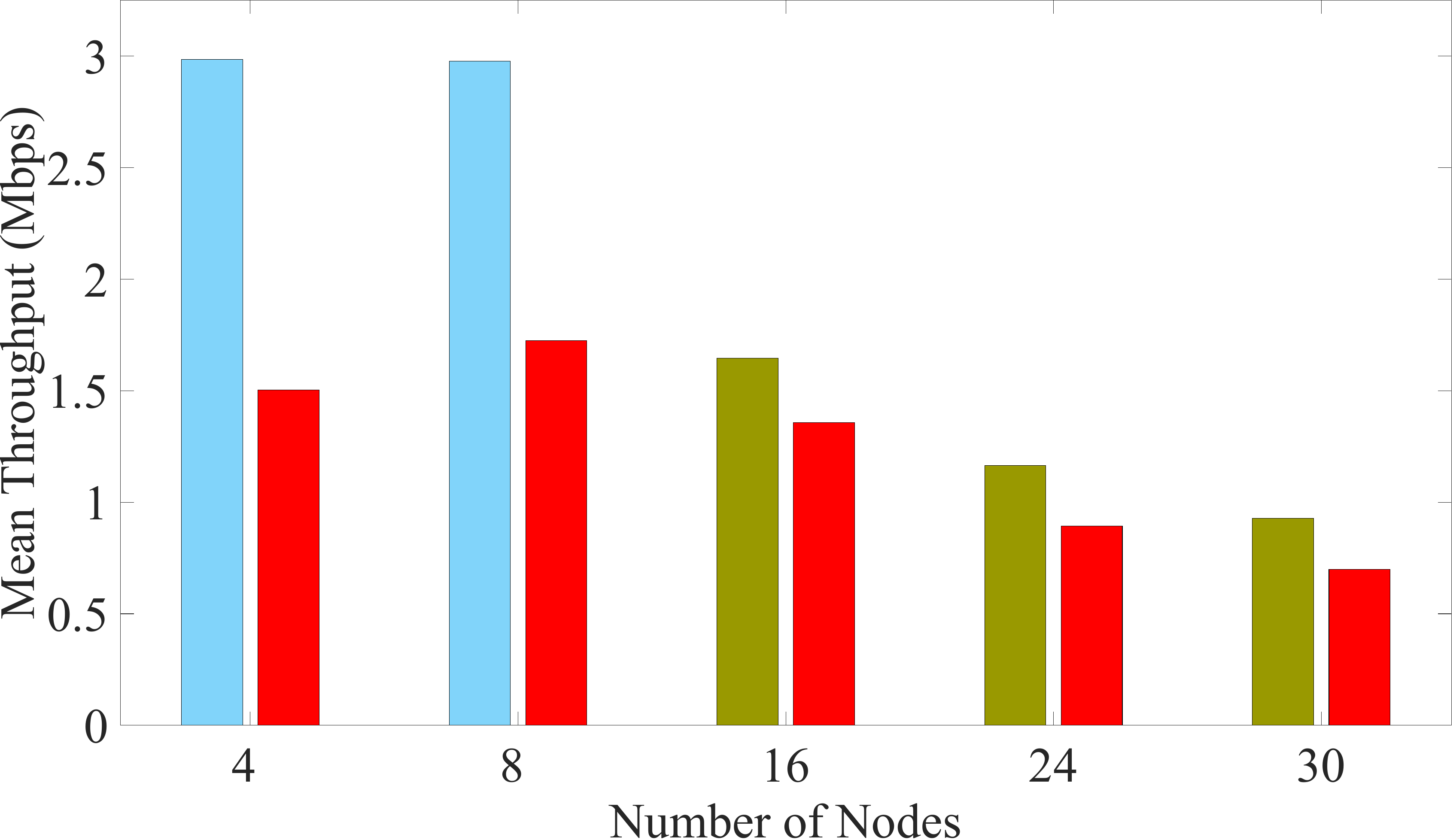}
		\label{TSUS_3}	
	}
  \addtocounter{subfigure}{-1}
 \subfigure[{250 packets/second average arrival rate}]{
		\includegraphics[width=0.315\textwidth,trim = 0mm 0mm 0mm 0mm,clip]{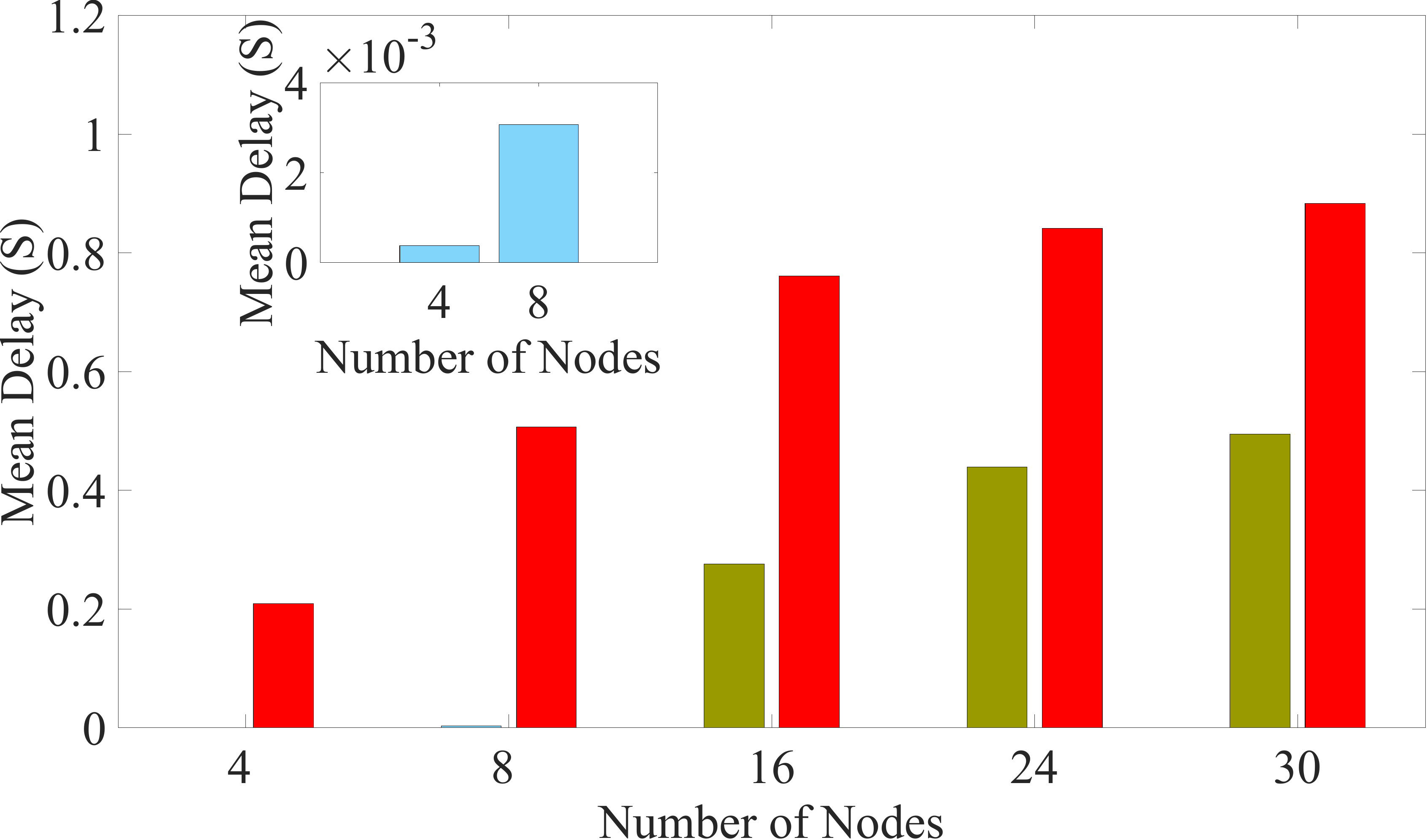}
		\label{TSUS_2_3}	
	}
 }
 }
	% \caption{Using Mean throughput as Reward}
	% \label{jain_index_2}
 % \end{figure*}
 % \begin{figure*}[!ht]
	% \centering
 %  \addtocounter{subfigure}{-3}
	% \subfigure[{20 packets/second average arrival rate}]{
	% 	\includegraphics[width=0.315\textwidth,trim = 0mm 0mm 0mm 0mm,clip]{Paper1/pic/Final/_fina_Mean_delay_rewardthroughput2.pdf}
	% 	\label{TSUS_2_1}	
	% }
%  \vrule
	
% \vrule
	
% \hline
\vspace{0.15cm}
\noindent\begin{minipage}{\linewidth}
\centering
%\small
%\footnotesize
\tiny
\begin{tblr}{
  hlines,
  vlines,
cell{2}{1}={c,red!},
  cell{3}{1}={c,Mycolor3},
  cell{4}{1}={c,Mycolor4},
  cell{5}{1}={c,Mycolor5},
  cell{6}{1}={c,Mycolor6},
  % cell{7}{1}={c,Mycolor7},
  cell{7}{1}={c,Mycolor8},
  cell{8}{1}={c,Mycolor12},
  cell{9}{1}={c,Mycolor9},
  cell{10}{1}={c,Mycolor10},
  %cell{11}{1}={c,Mycolor11},
  %cell{3}{1}={c,cyan}
  %cell{2}{1}={c,red!},
}
{\diagbox[width=11em]{\textbf{Protocol}}{\textbf{MAC blocks}}} & \textbf{Carrier Sensing } & {\textbf{Backoff}\\\textbf{  }} & {\textbf{Min CW }\\\textbf{ Size }} & \textbf{Slot Size } & \textbf{CTS/RTS } & \textbf{Data Rate } \\
\cellcolor{FD6864}\mbox{IEEE 802.11ac}                 & On                      & BEB                                              & $15$                                  & $9~\mu s$                                      & On                &  AARF Rate Control                 \\
{AI-Based Protocol (1)}                      & Off                   & ----                                       & ----                               & ----                              & Off             & $13~Mbps$            \\ 
 {AI-Based Protocol (2)}                          &       Off                     &     ----                                                    &                ----                     &          ----                                &        Off           &        $26~Mbps$              \\{AI-Based Protocol (3)} 
                          &    Off                       &        ----                                                &         ----                            &        ----                                  &         On          &     $26~Mbps$      \\{AI-Based Protocol (4)} 
                          &       On                    &    EIED                                                    &     15                                &              $5~\mu s$                           &        Off           &      $52~Mbps$    \\
                          % {AI-Based Protocol (5)} 
                          % &       On                    &    Off                                                    &     ----                               &              $5~\mu s$                           &        Off           &      $26~Mbps$     \\
                          {AI-Based Protocol (6)} 
                          &       On                    &    Constant                                                    &     15                               &              $5~\mu s$                           &        Off           &      $52~Mbps$
                          \\{AI-Based Protocol (7)} 
                          &       On                    &    Constant                                                    &     31                               &              $5~\mu s$                           &        Off           &      $52~Mbps$  
                          \\{AI-Based Protocol (8)} 
                          &   On                        &      EIED                                                   &            15                         &            $5~\mu s$                              &    Off               & $52~Mbps$          
                          \\{AI-Based Protocol (9)} 
                          &   On                        &      EIED                                                   &            $31$                         &            $5~\mu s$                              &    Off               & $78~Mbps$ 
                          % \\{AI-Based Protocol (9)} 
                          % &   On                        &      EIED                                                   &            $31$                         &            $5~\mu s$                              &    Off               & $78~Mbps$ 
\end{tblr}
\end{minipage}
% \end{table*}
 %\vspace{-0.10cm}
	\caption{Comparison of mean throughput and delay results with \mbox{IEEE 802.11ac}}
	\label{Evaluation_2}
 \vspace{-0.6cm}
\end{figure*}
\section{Performance Evaluation}
\vspace{-0.13cm}
We consider a layout of $150\times150$ $m^2$, with nodes distributed randomly across this area. The MAC policy is shared among all nodes to ensure a consistent policy execution across the entire network simultaneously.
%The designed MAC protocol was applied on all nodes.
Due to the use of DRL, one of the notable strengths of our framework is its adaptability to scenarios with diverse traffic patterns extending beyond those encountered during its training phase.
%To demonstrate the capability of our framework, which, due to using DRL, can even handle scenarios with different traffic 
%patterns than those used for training.

% We use the Poisson traffic pattern with a mean arrival packet number of $35$, $50$, $100$, $200$, and $350$ for training the agent.
For the training, we use Poisson traffic with a mean arrival packet rate ranging from 35 to 350 packets per second, as given in Table \ref{tab:Training Parameter}. The number of networks within the environment ranges from 1 to 30.
In this section, we analyze the performance of our AI-based protocol with respect to the average throughput of all networks and the mean end-to-end delay, and benchmark it against the \mbox{IEEE 802.11ac} protocol. \mbox{IEEE 802.11ac} employs carrier sensing, binary exponential backoff, a $9~\mu s$ slot time, and uses CTS/RTS mechanisms. Additionally, we enable adaptive auto rate fallback (AARF) as a rate controller mechanism for \mbox{IEEE 802.11ac}. 
%We use the 802.11 ac protocol as a baseline and compare the AI-based protocol outcomes to it.
% The 802.11 ac uses carrier sensing with binary exponential backoff and slot time of $9~\mu s$ with enabled CTS/RTS and adaptive rate control.

In Figure \ref{Evaluation_2}, we show the evaluation results from the AI-based protocol obtained across three distinct traffic patterns, each with varying numbers of coexisting parallel networks. We consider Poisson traffic with an average packet arrival rate of $20$, $150$, and $250$ packets per second. The number of networks in the environment varies from 4 to 30. Each subfigure presents the results for different traffic intensities. The first row depicts the achieved mean throughput, whereas the second row represents the mean end-to-end delay. The agent's decision varies according to the environmental characteristics, as indicated by the different colours. Each colour represents a combination of different MAC blocks, as illustrated in the accompanying table. The chosen MAC blocks represent the most frequently selected action by the agent in a single episode.

Figure \ref{TSUS_1} shows the mean throughput and delay of all networks for low-traffic scenarios.
The simulation results demonstrate a significant improvement in both mean throughput and mean delay. The mean throughput results demonstrate an approximate doubling of throughput performance compared to conventional \mbox{IEEE 802.11ac} protocols. Meanwhile, the mean delay in our AI-based protocol ranges from $40~\mu s$ to $80~\mu s$, whereas the \mbox{IEEE 802.11ac} MAC protocol shows delays on the order of $1~ms$ for scenarios with fewer coexisting networks. In overcrowded scenarios (24 and 30 nodes), the \mbox{IEEE 802.11ac} MAC protocol shows a delay in the order of $45~ms$ to $80~ms$.
% For enhanced visualization, we magnify the obtained delay outcome.
By analyzing the throughput and delay results, we observe that in scenarios with a low number of nodes, our AI-based protocol optimizes performance by deactivating the CS and CTS/RTS functions. This strategy is coupled with the use of low modulation and coding schemes to enhance the robustness to interference. Consequently, this decision results in higher mean throughput and reduced latency, primarily by eliminating the CS and CTS/RTS overhead.
As the number of networks in the environment increases to 16, our AI-based protocol keeps the CS and backoff functions in a deactivated while enabling CTS/RTS. This configuration ensures that neighbouring networks are aware of the channel occupancy status, achieved through the triggering of the network allocation function. In highly congested scenarios, our agent activates CS and implements EIED backoff. Additionally, the slot duration is reduced from $9~\mu s$ to $5~\mu s$, effectively mitigating the CS overhead. 
 
 In Figure \ref{TSUS_2}, we show the results for scenarios with medium traffic. The results clearly demonstrate the substantial performance advantages of the AI-based protocol in comparison to the \mbox{IEEE 802.11ac} protocol, particularly in scenarios with a lower number of nodes. Notably, the AI-based protocol achieved nearly double the average throughput of the \mbox{IEEE 802.11ac} protocol. Furthermore, it demonstrated significantly reduced latency when compared to the conventional \mbox{IEEE 802.11ac} protocol.
%The results indicate that the AI-based protocol achieved nearly double the average throughput of 802.11 ac for the scenarios with fewer nodes. Furthermore, the AI-based protocol exhibited a drastically decreased delay compared to the 802.11 ac protocol.
In environments with a lower number of nodes, the protocol activates CS while employing a constant backoff mechanism with a contention window size of $15$. In contrast, in overcrowded scenarios, the contention window size is adjusted to 31. In all medium traffic scenarios, CS remains active, but the slot duration is reduced to $5~\mu s$ to minimize CS overhead and enhance overall network performance.

Figure \ref{TSUS_3} depicts the results for a scenario with not-saturated high-load traffic. The results show that for scenarios with fewer nodes, we achieve almost twice the \mbox{IEEE 802.11ac} throughput with significantly reduced delay.
In these scenarios, we observed that the agent selects EIED backoff with a minimum contention window size of $15$. 
As the number of nodes increases, the AI-protocol still achieves almost the same throughput as the \mbox{IEEE 802.11ac} protocol but with a lower delay. In these scenarios, the agent chooses the EIED backoff block with a minimum contention window size of $31$. By employing the backoff procedure, the interference is reduced, and since the environment experiences high traffic, the agent protocol opts for higher transmission rates, leading to the observed improvements in throughput and lower delay compared to the \mbox{IEEE 802.11ac} protocol.

By analyzing the mean throughput curves in all three subfigures, we observe that our AI-based protocol outperforms the legacy \mbox{IEEE 802.11ac} protocol in all scenarios with different numbers of nodes and traffic. The resulting end-to-end delay shows that, in almost all scenarios, we get significantly less delay than the traditional protocol owing to the selection of the right data rate and the reduction of CS overheads. Another reason for the delay decrease is that the agent enhances the CS function by reducing the slot time to $5~\mu s$, which saves time during backoff and inter-frame space such as DIFS. By optimizing these time intervals, the agent decreases delays in the overall transmission process, leading to enhanced system efficiency and lower latency.
\section{Conclusions}
\vspace{-0.05cm}
In this paper, we proposed a framework that can be used to engineer adaptive and intelligent MAC protocols from components of legacy IEEE 802.11 by leveraging deep reinforcement learning techniques. Our results show that the AI-based protocol not only improves the mean network throughput, but also significantly reduces the latency in diverse scenarios. This research opens up exciting possibilities for the development of next-generation MAC protocols that can dynamically adjust their behavior in real-time that will be enhancing the QoS for diverse applications and services.
\vspace{-0.03cm}
\section*{Acknowledgment}
\vspace{-0.06cm}
\small
The authors acknowledge the financial support by the Federal Ministry of Education and Research of Germany in the project “Open6GHub” (grant number:16KISK012). 
% \section*{Acknowledgment}
% The authors acknowledge the financial support by the Federal Ministry of Education and Research of Germany in the project “Open6GHub” (grant number:16KISK012). Simulations were performed with computing resources granted by RWTH Aachen University under project rwth0767.
\vspace{-0.15cm}
\bibliographystyle{IEEEtran}
\bibliography{icc_radcom}

% Generated by IEEEtran.bst, version: 1.14 (2015/08/26)
\begin{thebibliography}{10}
\providecommand{\url}[1]{#1}
\csname url@samestyle\endcsname
\providecommand{\newblock}{\relax}
\providecommand{\bibinfo}[2]{#2}
\providecommand{\BIBentrySTDinterwordspacing}{\spaceskip=0pt\relax}
\providecommand{\BIBentryALTinterwordstretchfactor}{4}
\providecommand{\BIBentryALTinterwordspacing}{\spaceskip=\fontdimen2\font plus
\BIBentryALTinterwordstretchfactor\fontdimen3\font minus \fontdimen4\font\relax}
\providecommand{\BIBforeignlanguage}[2]{{%
\expandafter\ifx\csname l@#1\endcsname\relax
\typeout{** WARNING: IEEEtran.bst: No hyphenation pattern has been}%
\typeout{** loaded for the language `#1'. Using the pattern for}%
\typeout{** the default language instead.}%
\else
\language=\csname l@#1\endcsname
\fi
#2}}
\providecommand{\BIBdecl}{\relax}
\BIBdecl

\bibitem{Falko}
S.~Szott and et~al., ``{Wi-Fi Meets ML: A Survey on Improving IEEE 802.11 Performance With Machine Learning},'' \emph{IEEE Communications Surveys \& Tutorials}, vol.~24, no.~3, pp. 1843--1893, 2022.

\bibitem{giordano2023wifi}
L.~G. Giordano \emph{et~al.}, ``What will wi-fi 8 be? a primer on ieee 802.11bn ultra high reliability,'' \emph{arXiv preprint arXiv:2303.10442}, 2023.

\bibitem{Carlos}
E.~Reshef and C.~Cordeiro, ``{Future Directions for Wi-Fi 8 and Beyond},'' \emph{IEEE Commun. Mag.}, vol.~60, no.~10, pp. 50--55, 2022.

\bibitem{rate_Infocom}
S.-C. Chen \emph{et~al.}, ``An experience driven design for ieee 802.11ac rate adaptation based on reinforcement learning,'' in \emph{IEEE INFOCOM}, 2021, pp. 1--10.

\bibitem{Szymon_WCNC}
W.~Wydmański and S.~Szott, ``Contention window optimization in ieee 802.11ax networks with deep reinforcement learning,'' in \emph{2021 IEEE Wireless Commun. Netw. Conf.}, 2021, pp. 1--6.

\bibitem{Fawzi2022}
\BIBentryALTinterwordspacing
A.~Fawzi and et~al., ``Discovering faster matrix multiplication algorithms with reinforcement learning,'' \emph{Nature}, vol. 610, no. 7930, pp. 47--53, Oct 2022. [Online]. Available: \url{https://doi.org/10.1038/s41586-022-05172-4}
\BIBentrySTDinterwordspacing

\bibitem{Jakob1}
M.~P. Mota \emph{et~al.}, ``{The Emergence of Wireless MAC Protocols with Multi-Agent Reinforcement Learning},'' in \emph{Proc. IEEE GC Wkshps}, 2021, pp. 1--6.

\bibitem{Jean-Marie}
------, ``{Scalable Joint Learning of Wireless Multiple-Access Policies and their Signaling},'' in \emph{Proc. IEEE VTC2022-Spring}, 2022, pp. 1--5.

\bibitem{Peng}
P.~Wang \emph{et~al.}, ``{DMDL: A hierarchical approach to design, visualize, and implement MAC protocols},'' in \emph{Proc. IEEE WCNC}, 2018, pp. 1--6.

\bibitem{Pasandi}
H.~B. Pasandi and T.~Nadeem, ``Towards a learning-based framework for self-driving design of networking protocols,'' \emph{IEEE Access}, vol.~9, pp. 34\,829--34\,844, 2021.

\bibitem{Rec_mac}
[Online.], Available: \url{https://gitlab.com/navid-keshtiarast/Rec\_mac}.

\bibitem{802.11ac}
``{IEEE Standard for Information technology- Part 11: Wireless LAN Medium Access Control (MAC) and Physical Layer (PHY) Specifications},'' \emph{IEEE Std 802.11-2016 (Revision of IEEE Std 802.11-2012)}, pp. 1--3534, 2016.

\bibitem{omnet}
\BIBentryALTinterwordspacing
``Omnet++, discrete event simulator.'' [Online]. Available: \url{https://www.omnetpp.org/}
\BIBentrySTDinterwordspacing

\bibitem{INET}
\BIBentryALTinterwordspacing
``Inet framework.'' [Online]. Available: \url{https://inet.omnetpp.org/}
\BIBentrySTDinterwordspacing

\bibitem{schulman2017trust}
J.~Schulman \emph{et~al.}, ``Trust region policy optimization,'' \emph{arXiv preprint arXiv:1502.05477}, 2017.

\bibitem{schulman2017proximal}
------, ``Proximal policy optimization algorithms,'' \emph{arXiv preprint arXiv:1707.06347}, 2017.

\end{thebibliography}
\end{document}